

\magnification=1200
\voffset= -.2 truein
\vsize=7.3in
\hsize=5.4in
\def\no{\noindent}
\def\sy{\scriptscriptstyle}

\def\square
{\kern1pt\vbox{\hrule height
1.2pt\hbox{\vrule width 1.2pt\hskip 3pt
\vbox{\vskip 6pt}\hskip
3pt\vrule width 0.6pt}\hrule height 0.6pt}\kern1pt}

\def\O{{\cal O}}

\def\sy{\scriptscriptstyle}
\def\dps{\displaystyle}
\def\D12{{\tau_1 - \tau_2}}
\def\D13{{\tau_1 - \tau_3}}
\def\D23{{\tau_2 - \tau_3}}
\def\half{{1\over 2}}
\def\pa{\partial}

\def\g12{{\dot {G_B}_{12}}}
\def\g13{{\dot {G_B}_{13}}}
\def\g23{{\dot {G_B}_{23}}}
\def\lag{( -\partial^2 + V)}
\def\Tint{{\dps\int_{0}^{\infty}}{dT\over T}e^{-m^2T}}
\def\pint{{\dps\int}{dp_i\over {(2\pi)}^d}}
\def\Tr{{\rm Tr}\,}
\def\tr{{\rm tr}\,}
\def\del{\partial}
\def\deli{\partial_{\kappa}}
\def\delj{\partial_{\lambda}}
\def\delk{\partial_{\mu}}
\def\delij{\partial_{\kappa\lambda}}
\def\delik{\partial_{\kappa\mu}}
\def\deljk{\partial_{\lambda\mu}}
\def\delki{\partial_{\mu\kappa}}
\def\delkl{\partial_{\mu\nu}}
\def\delijk{\partial_{\kappa\lambda\mu}}
\def\deljkl{\partial_{\lambda\mu\nu}}
\def\delikl{\partial_{\kappa\mu\nu}}
\def\delijkl{\partial_{\kappa\lambda\mu\nu}}
\def\delijklm{\partial_{\kappa\lambda\mu\nu o}}

\def\g{\mbox{g}}

\def\O(#1){O($T^#1$)}

\def\eins{  1\!{\rm l}  }

\baselineskip 13pt plus 1pt minus 1pt
\begingroup\nopagenumbers
\hskip 10cm HD-THEP-93-44
\vskip 5pt
\centerline{{\bf THE HIGHER DERIVATIVE EXPANSION OF THE
EFFECTIVE ACTION}}
\centerline{{\bf  BY THE STRING--INSPIRED METHOD, PART I} }
\vskip 36pt
\centerline{{Denny Fliegner, Michael G. Schmidt and Christian Schubert
\footnote{*}{Partially supported by funds
provided by the Deutsche Forschungsgemeinschaft}
\footnote{**}{e-mail address schubert@hades.ifh.de}}}

\vskip 12pt

\centerline{{Institut f\"ur Theoretische Physik}}
\centerline{Universit\"at Heidelberg}
\centerline{Philosophenweg 16}
\centerline{D-69120 Heidelberg}
\vskip 10pt
\centerline{\bf December 1993}
\vskip 20pt
\centerline{\bf ABSTRACT}
{\medskip\narrower\noindent
The higher derivative expansion of the
one-loop effective action for an external scalar potential is
calculated to order O($T^7$), using the
string-inspired Bern-Kosower method in the first quantized path
integral formulation. Comparisons are made
with standard heat kernel calculations and with the corresponding
Feynman diagrammatic calculation in order to show the efficiency
of the present method.
\medskip}

\eject
\bigskip
\endgroup
\pageno = 1

As Bern and Kosower have shown [1], the possibility of
representing amplitudes in
ordinary quantum field theory as the infinite string tension limit of certain
superstring amplitudes allows the derivation of a new set of Feynman rules
for one-loop calculations which, though very different from the usual ones,
appear to be completely equivalent [2]. These rules
lead to a significant reduction of the number of terms to be computed in gauge
theory calculations, as they combine contributions of different Feynman
diagrams into gauge invariant structures, opening up the road to performing
a number of calculations beyond the reach of standard methods, such
as five-gluon-amplitudes [3] and four-graviton-amplitudes [4].
Strassler [5] later succeeded in deriving essentially the same set of rules
without explicit reference to string theory, using
path-integrals on the circle, and this reformulation of
the Bern-Kosower method  appears to be particularly well-suited to the
calculation of effective actions [6,7].

In this paper, we demonstrate
the efficiency of the method
by performing the -- to our knowledge first --
calculation to order O($T^7$)
of the higher derivative expansion of the effective action
for an external scalar potential induced by a scalar loop
(in this special case the method reduces in principle
to earlier work by
Polyakov [8]).
In spite of the
absence of gauge symmetry,
the string method turns out to hold significant
advantages over standard methods both regarding the
calculation of coefficients and the redundancy of the resulting operator
basis, a fact which we will try to elucidate
by comparing with several other types of heat kernel calculations.
The connection with Feynman diagram calculations will also be
clarified.

This special case of the higher derivative expansion is physically
interesting in cases where the gauge field degrees of freedom are
(at least in principle) already integrated out, or where they can be
eliminated in another way. The latter is the case in sphaleron
calculations in the 1+1-dimensional Abelian Higgs model [9],
where the one-loop fluctuation determinants are determinants of
P\"oschl-Teller-operators. They can be calculated exactly [10],
making the model an interesting playing ground for studying the
convergence of the higher derivative expansion, as a useful preparation
for the case of the electroweak sphaleron where the higher derivative is
essential as those nonapproximative results which are available
do contradict each
other [11].
The first case appears most
prominently in the discussion of radiative corrections to the bounce
configuration in the electroweak phase transition [12].
In this case an accuracy beyond the order O($T^6$) is highly
desirable in order to make the discussion more quantitative.

The application of the string method to the calculation of the
higher derivative expansion of the effective action with both a
scalar potential and a nonabelian gauge field
will be treated in
a separate publication [13].

The higher derivative expansion of an effective action
exists in two versions, which differ by the grouping of terms.
The one we will consider is commonly
obtained by writing the one-loop determinant
in the Schwinger proper time representation

$$\Gamma [V] = -  {\rm log}({\rm det}\,M) =
- {\rm Tr}({\rm log}\,M) = {\dps\int_0^{\infty}}{dT\over T}
{\rm Tr}\,e^{-TM}\eqno(1)$$

\no
and expanding in powers of $T$. For the case we will be treating in this
paper,

$$M = -\partial^2  + m^2 + V(x) \eqno(2)$$

\no
with $V(x)$ some matrix--valued function on spacetime, this expansion has been
previously
calculated to order O($T^6$) by various methods [9,14].

Alternatively, some authors prefer to calculate the same series up to
a fixed number of derivatives, but with an arbitrary number of fields
or potentials [15--17].

Let us first present the calculation by the string--inspired method, and
compare with other methods later. For the sake of
completeness, we will start from scratch and recapitulate how formula (1)
for the effective action may be transformed into a worldline path
integral (see e.g. [18]).

All calculations will be performed in $d$-dimensional
euclidean spacetime; operator traces will be denoted by $\Tr$, and
ordinary matrix traces by $\tr$.

Starting with the proper time integral

$$\Gamma\lbrack V\rbrack =
{\dps\int_{0}^{\infty}}{dT\over T}e^{-m^2T}\,
\Tr exp\bigl[-T\lag \bigr]\eqno(3)$$

\no
one evaluates the trace in $x$--space and decomposes the interval T in the
exponential, giving $(x_N = x_0 = x)$

$$\eqalign{
\Gamma [V] &= \Tint\tr{\dps\int}dx
<x\mid exp\bigl[-T\lag \bigr]\mid x>\cr
&= \Tint\tr
{\dps\int\prod_{i=1}^N}dx_i
<x_i\mid exp\biggl[-{T\over N}\lag\biggr]\mid x_{i-1}>\cr
&=  \Tint\tr{\dps\int\prod_{i=1}^N}dx_i K_{i,i-1}\cr
}\eqno(4)$$

\no
The kernel $K_{i,i-1}$ may be evaluated by insertion of a
momentum basis $\lbrace\mid p_i>\rbrace$,

$$\eqalign{
K_{i,i-1} & = \pint <x_i\mid exp\biggl[-{T\over N}(-{\del}^2 + V)\biggr]
            \mid p_i><p_i\mid x_{i-1}> \cr
      &= \pint \,exp\biggl[-{T\over N}\biggl(p_i^2+V(x_i)+ ip_i
        {(x_{i-1}-x_i)\over {T/N}}\biggr)\biggr]\cr
}\eqno(5)$$

\no
Taking $N$ to be large and performing the gaussian momentum integrations yields

$$K_{i,i-1} = {1\over {(4\pi {T/N})}^{d\over 2}}\, exp
     \biggl[-{T\over N} \bigl({{\dot x_i}^2\over 4}+ V(x_i)\bigr)\biggr],
\eqno(6)$$

\no
and in the limit of $N \to \infty$ one obtains a representation for $\Gamma$
as a quantum mechanical path integral on the space of closed paths in
spacetime:

$$\Gamma\lbrack V\rbrack =  {\dps\int_0^{\infty}}
{dT\over T}
{\lbrack 4\pi T\rbrack}^{-{d\over 2}}
e^{-m^2T}
\tr{\cal P}{\dps\int_{x(T)=x(0)}}\!\!\!\!\!\!\!\!\!\!\!\!\!\!\!\! {\cal D} x
\,exp\Bigl [- \int_0^T d\tau
\Bigl ({ \dot x^2\over 4}
+ V(x)
\Bigr )\Bigr ],
\eqno(7)$$

\no
where $\cal P$ denotes path ordering and
the normalization of the path integral is such that

$$\tr {\cal P}{\dps\int}
{\cal D}x\; exp\Bigl [- \int_0^T d\tau
\Bigl ({\dot x^2\over 4}
\Bigr )\Bigr ] = 1             \eqno(8)$$

\no
The existence of such path integral representations for
one-loop amplitudes is, of course,
nothing new in principle [19 -- 21]. The novel
feature of the string-inspired method consists rather in the mode of
their evaluation, which is by Wick contractions on the worldline, using
the
one-dimensional Green function for the Laplacian on the circle [8]:

$$\eqalign{
\langle x^{\mu}(\tau_1)x^{\nu}(\tau_2)\rangle
   &=  - g^{\mu\nu}G(\tau_1,\tau_2)
   = \quad - g^{\mu\nu}\biggl[ \mid \tau_1 - \tau_2\mid -
{{(\tau_1 - \tau_2)}^2\over T}\biggr]\cr
}\eqno(9)$$

\no
This Green function can, however, not be applied to the path integral as it
stands. Partial integration on the circle yields the identity

$$\int_0^T d\tau_2\, G(\tau_1,\tau_2)\ddot x(\tau_2) =
 2 x(\tau_1) - {2\over T} \int_0^T d\tau_2 \, x(\tau_2)\qquad,\eqno(10)$$

\no
where the second term should vanish, indicating that
one should really integrate over the relative coordinate only.
Let us therefore introduce the loop center of mass $x_0$,
writing

$$x^{\mu}(\tau) = x^{\mu}_0  +  y^{\mu} (\tau )\eqno(11)$$

\no
with

$$\int_0^T d\tau\,   y^{\mu} (\tau ) = 0,\eqno(12)$$

\no
and extract the integral over the center of mass from
the path integral:

$${\dps\int}{\cal D}x =
{\dps\int}d^d x_0{\dps\int}{\cal D} y.\eqno(13)$$

\no
Then Taylor-expanding the interaction part with respect to the center
of mass,

$$V(x(\tau )) = e^{y \partial}V(x_0),$$

\no
and expanding the path-ordered interaction exponential we get

$$\eqalign{
\Gamma\lbrack V\rbrack &= {\dps\int_0^{\infty}}
{dT\over T}{\lbrack 4\pi T\rbrack}^{-{d\over 2}}e^{-m^2T}
\tr{\dps\int}d^d x_0
{\dps\sum_{n=0}^{\infty}}\,{{(-1)}^n\over n}\,T
\int_0^{\tau_1 = T}d\tau_2\int_0^{\tau_{2}} d\tau_3
\ldots\int_0^{\tau_{n-1}} d\tau_n\cr
&\times
{\dps\int} {\cal D} y\,
e^{y({\scriptscriptstyle\tau_1})
\partial_{(1)}}V^{(1)}(x_0)
\ldots
e^{y({\scriptscriptstyle\tau_n})
\partial_{(n)}}V^{(n)}(x_0)
\,exp \Bigl [- \int_0^T d\tau
  \,{{\dot y}^2\over 4}\Bigr ].
}\eqno(14)$$

\no
We have labeled the background fields $V^{(1)},\ldots,V^{(n)}$,
and the first $\tau\,$ -- integration has
been eliminated by using the
freedom of choosing the point $0$ somewhere on the loop, which is also
the origin of the factor of ${1\over n}$.

The path integrals for fixed $n$ can now be performed using the
Wick contraction formula for exponentials well-known from string theory,

$$\eqalign{
\langle e^{y({\sy\tau_1})\partial_{(1)}}
e^{y({\sy\tau_2})\partial_{(2)}}\rangle &=
e^{{\scriptstyle -G}({\sy\tau_1},
{\sy\tau_2})\partial_{(1)}\partial_{(2)}}\cr
}\eqno(15)$$

\no
yielding

$$\eqalign{
{\dps\int} &{\cal D} y\,
e^{y({\scriptscriptstyle\tau_1})
\partial_{(1)}}V^{(1)}(x_0)
\ldots
e^{y({\scriptscriptstyle\tau_n})
\partial_{(n)}}V^{(n)}(x_0)
exp \Bigl [- \int_0^T d\tau \,
  {{\dot y}^2\over 4}\Bigr ]  =\cr
&exp\Bigl [-{\sum_{i<k}G_B({\tau_i},
{\tau_k})\pa_{(i)}\pa_{(k)}}\Bigr ]
 V^{(1)}\ldots V^{(n)}.\cr
}\eqno(16)$$

\no
Finally, by a rescaling

$$\tau_i = T\,u_i,\qquad i=1,\ldots,n\eqno(17)$$

\no
and using the scaling property of $G$

$$G(\tau_1,\tau_2) = T\,G(u_1,u_2)\eqno(18)$$

\no
we arrive at

$$\eqalign{
\Gamma\lbrack V\rbrack &= {\dps\int_0^{\infty}}
{dT\over T}{\lbrack 4\pi T\rbrack}^{-{d\over 2}}e^{-m^2T}
\tr{\dps\int}d^d x_0\,
{\dps\sum_{n=0}^{\infty}}\,{{(-T)}^n\over n}\cr
&\times
\int_0^{u_1 = 1}du_2\int_0^{u_{2}} du_3
\ldots\int_0^{u_{n-1}} du_n\,
exp\Bigl [-T{\sum_{i<k}G_B({u_i},{u_k})\pa_{(i)}\pa_{(k)}}\Bigr ]
V^{(1)}\ldots V^{(n)}.
}\eqno(19)$$

\vfill\eject

\no

The higher derivative expansion to some fixed order in $T$ can now be simply
obtained by expanding the exponential, performing a number of
multi-integrals with polynomial integrands, which can be easily done with
MAPLE or MATHEMATICA,
and identifying operators which differ only by the
cyclic ordering under the trace.

\no
To order O($T^7$) this
can be done without excessive hardship, and leads to the following result:

$$\Gamma\lbrack V\rbrack = {\dps\int_0^{\infty}}
{dT\over T}{\lbrack 4\pi T\rbrack}^{-{d\over 2}}e^{-m^2T}\,
{\dps\sum_{n=1}^{7}}\,{(-T)}^n \tr O_n\, ,\eqno(20)$$

\vskip4pt
\no
where, using some shorthand-notation ($\delij V:= \deli\delj V(x_0)$ etc.) and
changing $x_0$ to $x$,

$$\eqalign{
O_1 &= \int dx \biggl(  V \biggr) \cr
\noalign{\vskip8pt}
O_2 &= {1\over 2!} \int dx \biggl( V^2 \biggr) \cr
\noalign{\vskip8pt}
O_3 &= {1\over 3!} \int dx
\biggl(  V^3
        + {1\over 2} \deli V \deli V
\biggr) \cr
\noalign{\vskip8pt}
O_4 &= {1\over 4!} \int dx
\biggl( V^4
        + 2 V \deli V \deli V
        + {1\over5} \delij V \delij V \biggr) \cr
\noalign{\vskip8pt}
O_5 &= {1\over 5!} \int dx \biggl(
        V^5
       + 3 V^2 \deli V \deli V
       + 2 V \deli V V \deli V
       + V \delij V \delij V
       + {5\over 3} \deli V\delj  V \delij V \cr
&\qquad\qquad       + {1\over 14} \delijk V \delijk V
\biggr)\cr
\noalign{\vskip8pt}
 O_6 &= {1\over6!} \int dx
  \biggl( V^6
      + 4 V^3 \deli V \deli V
      + 6 V^2 \deli V V \deli V
      + {12\over7} V^2 \delij V \delij V\cr
&\qquad\qquad      + {9\over7} V \delij V V \delij V
     + {26\over7} V \delij V \deli V \delj V
  + {26\over7} V \deli V \delj V \delij V\cr
&\qquad\qquad      + {17\over14} \deli V \delj V \deli V \delj V
     + {18\over7} V \deli V \delij V \delj V
      + {9\over7} \deli V \deli V \delj V \delj V\cr
&\qquad\qquad   + {3\over7} V \delijk V \delijk V
      + \delk V \delij V \delijk V
      + \delk V \delijk V \delij V\cr
&\qquad\qquad      + {11\over21} \delij V \deljk V \delki V
      +  {1\over42} \delijkl V \delijkl V \biggr)\cr
}$$

\vfill\eject

$$\eqalign{
 O_7 &= {1\over7!} \int dx
  \biggl(  V^7
   + 5 V^4 \deli V \deli V
     + 8 V^3 \deli V V \deli V
     + {9\over2} V^2 \deli V V^2 \deli V
     + {5\over2} V^3 \delij V \delij V  \cr
&\quad+ {9\over2} V^2 \delij V V \delij V
     + 6 V^2 \delij V \deli V \delj V
     + 6 V^2 \deli V \delj V \delij V
     + {7\over2} V^2 \deli V \delij V \delj V \cr
&\quad+ {17\over2} V \deli V \delj V \deli V \delj V
     + {7\over2} V \deli V \deli V \delj V \delj V
     + {11\over2} V \deli V \delj V \delj V \deli V\cr
&\quad     + {11\over2} V \deli V V \delij V \delj V
 + {11\over2} V \deli V V \delj V \delij V
     +{17\over2} V \delij V V \deli V \delj V\cr
&\quad     +{2\over3} V \delijk V V \delijk V
     +{5\over6} V^2 \delijk V \delijk V
  + {5\over2} V \delijk V \delij V \delk V\cr
&\quad     +{17\over6} V \delijk V \deli V \deljk V
     +{5\over3} V \delij V \delijk V \delk V
     + {17\over6} V \delij V \delk V \delijk V\cr
&\quad  + {5\over2} V \deli V \deljk V \delijk V
     + {5\over3} V \deli V \delijk V \deljk V
     + {5\over3} \deli V \deli V \deljk V \deljk V\cr
&\quad     + {11\over6} \deli V \deljk V \deli V \deljk V
  + {35\over9} \delijk V \deli V \delj V \delk V
     + {11\over3} V \delij V \delik V \deljk V\cr
&\quad     + {35\over18} \delij V \deli V \deljk V \delk V
     + {35\over18} \delij V \delk V \deljk V \deli V
  + {97\over18} \delij V \delik V \delj V \delk V\cr
&\quad     + {43\over18} \delij V \delik V \delk V \delj V
     + {1\over6} V \delijkl V \delijkl V
     + {1\over2} \deli V \deljkl V \delijkl V\cr
&\quad   + {1\over2} \deli V \delijkl V \deljkl V
     + {7\over10} \delij V \delkl V \delijkl V
     + {16\over15} \delij V \delikl V \deljkl V\cr
&\quad     + {1\over132} \delijklm V \delijklm V \biggr)\cr
}$$

\no
We have not performed the final $T$--integration, as it simply yields a
$\Gamma$--function at any fixed order of $T$, which will have poles for the
part of the effective action which is going to be renormalized at one loop
for the dimension of spacetime considered.

The following two points should be emphasized about this calculation:

\no
i) The final result has been obtained without performing any partial
integrations with respect to $x$.

\no
ii) Cyclic invariance has not been broken, despite of the formal breaking of
cyclicity implied by distinguishing the variable $u_1$.

The book-keeping has still been done by hand,
though computerization is in progress,
and should allow calculation to several more orders in the
expansion.

In the commutative case our result may be written in a much
simpler form as, with $x_0$ fixed, $V(x_0)$ may be extracted from the
path integral in eq. (7):

\vfill\eject

$$\Gamma\lbrack V\rbrack = {\dps\int_0^{\infty}}
{dT\over T}{\lbrack 4\pi T\rbrack}^{-{d\over 2}}
\,\int dx
\,e^{-[m^2+V(x)] T}
\,{\dps\sum_{n=0}^{7}}\,{(-T)}^n \tilde O_n \, ,
\eqno(21)$$

$$\eqalign{
\tilde O_0 &= 1\cr
\noalign{\vskip5pt}
\tilde O_1 &= 0\cr
\noalign{\vskip5pt}
\tilde O_2 &= 0\cr
\noalign{\vskip5pt}
\tilde O_3 &= {1\over 3!}  \biggl(  {1\over 2} \deli V \deli V
\biggr) \cr
\noalign{\vskip5pt}
\tilde O_4 &= {1\over 4!} \biggl( {1\over5} \delij V \delij V \biggr) \cr
\noalign{\vskip5pt}
\tilde O_5 &= {1\over 5!} \biggl(
     {5\over 3} \deli V\delj  V \delij V
       + {1\over 14} \delijk V \delijk V
\biggr)\cr
\noalign{\vskip5pt}
\tilde  O_6 &= {1\over6!}   \biggl(
       {5\over2} ({\deli V  \deli V)}^2
      + 2 \delk V \delij V \delijk V
      + {11\over21} \delij V \deljk V \delki V\cr
&\qquad\qquad
      +  {1\over42} \delijkl V \delijkl V \biggr)\cr
\noalign{\vskip5pt}
\tilde O_7 &= {1\over7!}   \biggl(
     {7\over2} \deli V \deli V \deljk V \deljk V
  + {35\over9} \delijk V \deli V \delj V \delk V\cr
&\qquad\qquad
     + {35\over3} \delij V \deli V \deljk V \delk V
     +  \deli V \deljkl V \delijkl V
     + {7\over10} \delij V \delkl V \delijkl V\cr
&\qquad\qquad     + {16\over15} \delij V \delikl V \deljkl V
     + {1\over132} \delijklm V \delijklm V \biggr)\cr
}$$

\no

Now let us compare with the heat-kernel calculations of [9,14],
who have obtained the order O($T^6$) of this expansion (see also
[22] and references therein).

Belkov et al. [14] have been using the standard recursive $x$--space
calculation
of the heat-kernel coefficients for the operator $M$, based on the
separation ansatz

$$<x\mid K(T) \mid y>\quad = \quad <x\mid K_0(T)
\mid y>H(x,y;T)\eqno(22)$$

\no
where

$$\eqalign{
K&=exp\bigl[-T(-\del^2 + m^2 + V(x))\bigr]\cr
K_0&=exp\bigl[-T(-\del^2 + m^2)\bigr].
}\eqno(23)$$

\no
Inserted into the heat-kernel equation

$${\pa\over\pa T}K + MK = 0 \eqno(24)$$

\no
with boundary condition

$$K(T=0) = \eins \eqno(25)$$

\no
and using the known heat kernel of $M_0$ this
yields a recursive relation for the coefficients of the expansion of
$H$ in
powers of $T$,

$$H(x,y;T) = \sum_{k=0}^{\infty}\,a_k(x,y)\,T^k \eqno(26)$$

\no
This recursion process is too cumbersome to do without computer help, and not
easy to program for $d>1$ (on the other hand, for $d=1$ programming is trivial
and calculation easy up to O($T^{12}$)). More seriously, the expressions
for $O_6$ obtained by those authors are considerably more lengthy than our
result for the same quantity, and we have not been able to find the
necessary partial integrations to reduce them to our result.

Carson [9] (see also [23]) uses a
nonrecursive variant of the heat-kernel method first proposed by
Ne\-po\-me\-chie [24], using an insertion of plane waves into the functional
trace:

$$\Tr e^{-TM} = \tr\int dx{{\dps\int}{dp\over
{(2\pi)}^d}} e^{-ipx}e^{-TM}e^{ipx}
              = e^{-Tm^2}\tr\int dx
{{\dps\int}{dp\over {(2\pi)}^d}}
e^{-Tp^2}e^{-T\tilde M}\eins
\eqno(27)$$

\no
with

$$\tilde M = -\del^2 + V(x) -2ip\del\eqno(28)$$

\no
Now one expands $e^{-T\tilde M}\eins$ in $T$, and performs the momentum
integration, usually after a rescaling of the momenta by ${1\over\sqrt T}$
to avoid mixing of $T$-- orders by the integration.
As the final step, all derivatives have to be commuted to the right till
they vanish by acting on the $\eins$.

The main drawback of this procedure, at least for the case at hand,
is to be seen in the creation of an
enormous number of terms at intermediate stages of the calculations,
which cancel in the end.
Up to order O($T^5$) we have verified that this calculation leads to an
effective action which is related to our result by partial integrations.
At order O($T^6$) the result given by Carson is similarly compact as
ours, however this is after a considerable number of partial integrations
have been performed, as we have checked by redoing this calculation.
For better comparison,
we have pursued the calculation to order O($T^7$) by this method,
too, using FORM [25].
We needed 2 hours of CPU--time on a NEXT work station
to arrive at an O($T^7$)-part which
contains 155 terms before identification of cyclically permuted operators,
compared with 143 arising with the string method. Those 143 terms, however,
may be immediately reduced to the 37 of our
final result, due to the fact that
cyclic invariance has not been broken. This is not true
for the heat-kernel calculation, where this step would already be extremely
tedious to do by hand, and difficult to implement with FORM.

\no
For the special case of the one-dimensional P\"oschl-Teller potential

$$V_s(x) = -{s(s+1)\over cosh^2x},\eqno(29)$$

\no
relevant for the 1+1--dimensional Abelian Higgs model, we have checked
that the effective actions calculated to O($T^7$) by all three methods
do agree.

Much closer related to our calculation is the
earlier one by Zuk [15], who obtained
those terms in the O($T^6$) expansion which do not contain more than four
derivatives. This author
used Onofri's graphical method [26] for the heat kernel
expansion to arrive at the following expression for the $k$-derivative
contribution to the one-loop effective lagrangian (for the case of a
massless complex scalar):

$$\eqalign{
{\cal L}_k^1 &= - {\dps\int_0^{\infty}}{dT\over T}
{(4\pi T)}^{-{d\over 2}}\tr f^{(k)}(x;T),\cr
f^{(k)}(x;T) &= {\dps \sum_{n=1}^k}f_n^{(k)}(x;T),\cr
f_n^{(k)}(x;T) &= 2^{k\over 2}{(-1)}^n
{\dps\sum_{\alpha\in{\cal G}(n,k)}}w_{\alpha}
\int_0^{T}d\tau_1\int_0^{\tau_{1}} d\tau_2
\ldots\int_0^{\tau_{n-1}} d\tau_n
\,Q_{\alpha}(\tau ;T)\cr
&\times \Bigl[{\dps\prod_{i=1}^n}{1\over {r_i^{\alpha}!}}
e^{-(\tau_{i-1}-\tau_i)V}
\del_{\mu\kappa\lambda\dots}V^{(i)}\Bigr]
e^{-\tau_n V}\cr
}\eqno(30)$$

\no
Here  ${\cal G}(n,k)$  is the set of ordered graphs corresponding to all ways
of connecting $n$ points on a line by links in such a way that no point remains
unlinked, admitting self-links. To every link one associates a factor of

$$Q(\tau_i,\tau_j) \equiv \tau_j(1-{\tau_i\over T}) \eqno(31)$$

\no
and a contracted pair of derivatives acting on $V^{(i)},V^{(j)}$.
$Q_{\alpha}$ is the product of all $Q_{ij}$ for a graph,
and $w_{\alpha},{1\over {r_i^{\alpha}!}}$
are combinatorial weight factors .

This is obviously very similar to our eq. (19). And indeed, we may obtain
eq. (30) from eq. (19) by observing that

$$G_B(u_i,u_j) = Q(u_i,u_i) + Q(u_j,u_j) -2 Q(u_i,u_j)
\qquad\qquad (u_i\geq u_j),
\eqno(32)$$

\no
and using the relation

$${\dps\sum_{i=1}^n}\, \del_{(i)} = 0 \eqno(33)$$

\no
(valid up to total derivatives when acting on $V^{(1)}\ldots V^{(n)}$)
to rewrite

$$\sum_{i<k}G_B({u_i},{u_k})\pa_{(i)}\pa_{(k)}=
-\sum_{i=1}^n Q(u_i,u_i)\pa_{(i)}^2
-2\sum_{i<k} Q(u_i,u_j)\pa_{(i)}\pa_{(j)} \eqno(34)$$

\no
However, this transformation creates a huge redundancy; for instance, our
single term ${1\over 14} \delijk V \delijk V$ from $O_5$ splits up into
eight equivalent terms under it.

Finally, we comment on the calculation of the higher derivative
expansion by Feynman diagrams [27],
restricting to the non-matrix case for simplicity. Let us thus look at a
one-component scalar field theory with self-interaction $U(\phi)$,

$${\cal L} = \half (\pa_{\mu}\phi)^2 -
\half m^2 {\phi}^2 - U(\phi) \eqno(35)$$

\no
The euclidean one-loop effective action may be expressed as (see e.g. [28])

$$\Gamma(\phi) = - \Tr {\rm log}
\bigl[ -\del^2 + m^2 +  U''(\phi)\bigr] \eqno(36)$$

\no
and thus corresponds to the special case

$$U''\circ \phi = V\eqno(37)$$

\no
in our previous notation. To calculate the higher derivative expansion
up to order O($T^n$), one would have to calculate the one-loop diagrams
with up to $n$ insertions of the vertex $U''(\phi)$, and Taylor-expand
the diagram with $k$ vertices to order $2(n-k)$ in the external
momenta.
Using Feynman-parameters, this leads to formulas similar to
equation (19); for instance, in the case of the four-point diagram (choosing
$U(\phi) = {\lambda\over 4!}{\phi}^4$ for definiteness)
\vskip6cm

\no
after performing the momentum integrations one obtains an
$\alpha$--parameter representation

\vfill\eject

$$\Gamma_4\lbrack V\rbrack \sim
{\dps\int_{0}^{\infty}}{dT\over T}{[4\pi T]}^{-{d\over 2}}e^{-m^2T}
\,{\dps\prod_{i=1}^{4}}d\alpha_1\ldots d\alpha_4
\,\delta (T - \alpha_1 - \alpha_2 - \alpha_3 - \alpha_4)
\,{exp\bigl[-Q(p_i,\alpha_i)\bigr]},\eqno(38)$$

\no
where

$$
TQ = \alpha_2\alpha_4{(p_1+p_2)}^2 +
\alpha_1\alpha_3{(p_2+p_3)}^2 +
\alpha_1\alpha_4 p_1^2 +
\alpha_1\alpha_2 p_2^2 +
\alpha_2\alpha_3 p_3^2 +
\alpha_3\alpha_4 p_4^2
\eqno(39)$$

\no
Using the transformation of variables

$$\eqalign{
\alpha_1 &= T(u_1-u_2)\cr
\alpha_2 &= T(u_2-u_3)\cr
\alpha_3 &=  T(u_3-u_4)\cr
\alpha_4 &= T - (\alpha_1+\alpha_2+\alpha_3) =  T\bigl[1-(u_1-u_4)\bigr]\cr
}\eqno(40)$$

\no
and momentum conservation

$$p_1+p_2+p_3+p_4 = 0 \eqno(41)$$

\no
the exponent may be easily transformed into the Koba-Nielsen form

$$T\sum_{i<j}p_ip_j\,G_{ij} = T\Bigl\lbrace
p_1p_2\,\bigl[(u_1-u_2)-{(u_1-u_2)}^2\bigr]+
p_1p_3\,\bigl[(u_1-u_3)-{(u_1-u_3)}^2\bigr]+ \ldots\Bigr\rbrace
\eqno(42)$$

\no
(this type of transformation has been dubbed the
``stringy rearrangement'' by
Lam [29]).
However, we believe that this rearrangement, which restores the
symmetry among the external legs, is quite
essential for obtaining a relatively compact result for the effective
action at high orders.

Finally, we would like to caution that this compactness of
 our form for the effective action might be deceptive
with regard to applications requiring the use of
the equations
of motion, as it uses the mixed derivatives only, and one would have to perform
a large number of partial integrations to create box operators for all external
legs. On the other hand, it is easy to see that only up to order O($T^4$) the
effective action may be written in terms of box operators alone; beginning with
O($T^5$) the introduction of mixed derivatives is unavoidable.

\vskip.5cm
\no
We would like to thank N. Dragon, O. Lechtenfeld, M. Reuter, A. A. Tseytlin,
A. van de Ven and G. Veneziano for various helpful discussions and
informations.
C. S. would also like to thank the Department of Physics of Hannover
University for hospitality, and the
Deutsche Forschungsgemeinschaft for financial
support during part of this work.
\vfill\eject
\vskip1cm
\hskip3cm

{\bf References}

\vskip.5cm
\parindent=17pt

\item{[1]}{Z. Bern and D. A. Kosower,
Phys. Rev. Lett. {\bf 66} (1991) 1669;\hfill\break
Z. Bern and D. A. Kosower, Nucl. Phys. {\bf B379} (1992) 451;\hfill\break
Z. Bern, UCLA-93-TEP-5.}
\item{[2]}{Z. Bern and D. C. Dunbar, Nucl. Phys. {\bf B379} (1992) 562.}
\item{[3]}{Z. Bern, L. Dixon and D. A. Kosower, SLAC-PUB-5947;\hfill\break
           Z. Bern, L. Dixon and D. A. Kosower, UCLA-93-TEP-40.}
\item{[4]} {Z. Bern, D. C. Dunbar and T. Shimada, Phys. Lett. {\bf B312}
           (1993) 277.}
\item{[5]}{M. J. Strassler, Nucl. Phys. {\bf B385} (1992) 145.}
\item{[6]}{M. J. Strassler, SLAC-PUB-5978 (1992).}
\item{[7]}{M. G. Schmidt and C. Schubert, Phys. Lett. {\bf B318} (1993) 438.}
\item{[8]}{A. M. Polyakov, {\sl Gauge Fields and Strings}, Harwood 1987.}
\item{[9]} {L. Carson, Phys. Rev. {\bf D42} (1990) 2853.}
\item{[10]}{A. I. Bochkarev and G. G. Tsitsishvili, Phys. Rev. {\bf D40}
     (1989) 1378.}
\item{[11]} {L. Carson, X. Li, L. McLerran and R. Wang, Phys. Rev. {\bf 42 D}
    (1990) 2127;\hfill\break
     J. Baacke and S. Junker, DO-TH-93-19.}
\item{[12]} {J. Kripfganz, A. Laser and M. G. Schmidt, in preparation.}
\item{[13]} {D. Fliegner, M. G. Schmidt and C. Schubert, in preparation.}
\item{[14]} {A. A. Belkov, D. Ebert, A. V. Lanyov and A. Schaale,
      Preprint DESY-Zeuthen (June 1993).}
\item{[15]}{J. A. Zuk, Nucl. Phys. {\bf B280} (1987) 125.}
\item{[16]}{L. H. Chan, Phys. Rev. Lett. {\bf 57} (1986) 1199.}
\item{[17]}{J. Caro and L. L. Salcedo, Phys. Lett. {\bf B 309} (1993) 359.}
\item{[18]} {S. Iso, C. Itoi and H. Mukaida, Nucl. Phys. {\bf B346}
(1990) 293.}
\item{[19]}{R. P. Feynman, Phys. Rev. {\bf 80} (1950) 440.}
\item{[20]}{E. S. Fradkin, Nucl. Phys. {\bf 76} (1966) 588;
     C. J. Isham in {\sl Quantum Field Theory and Quantum Statistics},
I. Batalin, C. J. Isham and G. A. Vilkoviskii eds., Bristol and Hilger 1987.}
\item{[21]}{L. Alvarez-Gaum\'e, Commun. Math. Phys. {\bf 90} (1983) 161.}
\item{[22]}{R. D. Ball, Phys. Rep. {\bf 182} (1989) 1.}
\item{[23]} {L. Carson, L. McLerran, Phys. Rev. {\bf D41} (1990) 647.}
\item{[24]}{R. I. Nepomechie, Phys. Rev. {\bf D31} (1985) 3291.}
\item{[25]}{J. A. M. Vermaseren, FORM Manual, version 1, 1991.}
\item{[26]}{E. Onofri, Am. J. Phys. {\bf 46} (1978) 379.}
\item{[27]}{K. Kikkawa, Prog. Theor. Phys. {\bf 56} (1976) 947;
   \hfill\break
   R. MacKenzie, F. Wilczek and A. Zee, Phys. Rev. Lett. {\bf 53} (1984) 2203;
\hfill\break
C. M. Fraser, Z. Physik {\bf 28} (1985) 101.}
\item{[28]}{C. Itzykson and J. Zuber, {\sl Quantum Field Theory},
          McGraw-Hill 1980.}
\item{[29]}{C. S. Lam, MCGILL-93-20.}

\bye